 \documentclass[preprint2]{aastex}
\lefthead{Dewangan et al.}

\slugcomment{To apear in ApJ}

\newcommand\asca{{\it ASCA}}

\newcommand\chandra{{\it Chandra}}
\newcommand\rosat{{\it ROSAT}}

\newcommand\xmm{{\it XMM-Newton}}

\newcommand\s{{\rm~s}}
\newcommand\ks{{\rm~ks}}

\newcommand\hz{{\rm~Hz}}
\newcommand\kev{{\rm~keV}}
\newcommand\ev{{\rm~eV}}
\newcommand\kms{\ifmmode {\rm~km\ s}^{-1} \else ~km s$^{-1}$\fi}
\newcommand\Hunit{\ifmmode {\rm~km\ s}^{-1}\ {\rm Mpc}^{-1}
        \else ~km s$^{-1}$ Mpc$^{-1}$\fi}
\newcommand\ctssec{\ifmmode {\rm~count\ s}^{-1} \else ~count s$^{-1}$\fi}
\newcommand\ergsec{\ifmmode {\rm~erg\ s}^{-1} \else
        ~erg s$^{-1}$\fi}
\newcommand\funit{\ifmmode {\rm~erg\ s}^{-1}\;{\rm cm}^{-2} \else
        ~ergs s$^{-1}$ cm$^{-2}$\fi}
\newcommand\phflux{\ifmmode {\rm~photon\ s}^{-1}\;{\rm cm}^{-2}
        \else   ~photon s$^{-1}$ cm$^{-2}$\fi}
\newcommand\efluxA{\ifmmode {\rm~erg\ s}^{-1}\;{\rm cm}^{-2}\;{\rm
        \AA}^{-1} \else ~erg s$^{-1}$ cm$^{-2}$ \AA$^{-1}$\fi}
\newcommand\efluxHz{\ifmmode {\rm~erg\ s}^{-1}\;{\rm cm}^{-2}\;{\rm
        Hz}^{-1} \else ~erg s$^{-1}$ cm$^{-2}$ Hz$^{-1}$\fi}
\newcommand\cc{\ifmmode {\rm~cm}^{-3} \else cm$^{-3}$\fi}
\newcommand\FWHM{\ifmmode {\rm~FWHM} \else ${\rm~FWHM}$\fi}
\newcommand\Msun{\ifmmode M_{\odot} \else $M_{\odot}$\fi}
\newcommand\Lsun{\ifmmode L_{\odot} \else $L_{\odot}$\fi}
\newcommand\ltsim{\raisebox{-.5ex}{$\;\stackrel{<}{\sim}\;$}}

\newcommand\hbeta{\ifmmode {\rm H}\beta \else H$\beta$\fi}
\newcommand\Kalpha{\ifmmode {\rm K}\alpha \else K$\alpha$\fi}
\newcommand\nh{\ifmmode N_{\rm H} \else N$_{\rm H}$\fi}
\usepackage{graphicx}
\usepackage{here}

\begin{document}


\title{An investigation of the origin of soft X-ray excess emission
  from Narrow-line Seyfert 1 galaxies Akn~564 and Mrk~1044}

\author{G. C. Dewangan\altaffilmark{1}, R. E.
  Griffiths\altaffilmark{1}, Surajit Dasgupta\altaffilmark{2} \& A. R.
  Rao\altaffilmark{2}} \altaffiltext{1}{Department of Physics,
  Carnegie Mellon University, 5000 Forbes Avenue, Pittsburgh, PA 15213
  USA; {\tt email: gulabd@cmu.edu; griffith@seren.phys.cmu.edu} }
\altaffiltext{2}{Department of Astronomy \& Astrophysics, Tata
  Institute of Fundamental Research, Mumbai, 400005 India; {\tt email:
    surajit@tifr.res.in; arrao@tifr.res.in}}

\begin{abstract}
  We investigate the origin of the soft X-ray excess emission from
  narrow-line Seyfert 1 galaxies Akn~564 and Mrk~1044 using \xmm{}
  observations.  We find clear evidence for time delays between the
  soft and hard X-ray emission from Akn~564 based on a $\sim 100\ks$
  long observation.  The variations in the $4-10\kev$ band lag behind
  that in the $0.2-0.5\kev$ band by $1768\pm122\s$.  The full band
  power density spectrum (PDS) of Akn~564 has a break at $\sim
  1.2\times10^{-3}\hz$ with power-law indices of $\sim 1$ and $\sim 3$
  below and above the break. The hard ($3-10\kev$) band PDS is
  stronger and flatter than that in the soft ($0.2-0.5\kev$) band.  Based
  on a short observation of Mrk~1044, we find no correlation between
  the $0.2-0.3\kev$ and $5-10\kev$ bands at zero lag.  These
  observations imply that the soft excess is not the reprocessed hard
  X-ray emission. The high resolution spectrum of Akn~564 obtained
  with the reflection grating spectrometer (RGS) shows evidence for a
  highly ionized and another weakly ionized warm absorber medium. The
  smeared wind and blurred ionized reflection models do not describe
  the EPIC-pn data adequately.  The spectrum is consistent with a
  complex model consisting of optically thick Comptonization in a cool
  plasma for the soft excess and a steep power-law, modified by two
  warm absorber media as inferred from the RGS data and the foreground
  Galactic absorption.  The smeared wind and optically thick
  Comptonization models both describe the spectrum of Mrk~1044
  satisfactorily, but the ionized reflection model requires extreme
  parameters. The data suggest two component corona -- a cool,
  optically thick corona for the soft excess and a hot corona for the
  power-law component. The existence of a break in the soft band PDS
  suggests a compact cool corona that can either be an ionized surface
  of the inner disk or an inner optically thick region coupled to a
  truncated disk. The steep power-law component is likely arising from
  an extended region.

\end{abstract}

\keywords{accretion, accretion disks -- galaxies: active -- X-rays:
  galaxies}

\section{Introduction}
A significant fraction of type 1 active galactic nuclei (AGN) show
`soft X-ray excess emission' above a power-law continuum, usually
identified as the steepening of the X-ray continuum below $\sim
2\kev$. This soft excess emission was first observed by {\it HEAO-1}
(Pravdo et al. 1981), and {\it EXOSAT} (Arnaud et al. 1985; Singh et
al. 1985).  Boller, Brandt, \& Fink (1996) showed that AGN with the
steepest soft X-ray spectra tend to lie at the lower end of the
H$\beta$ line width distribution.  These AGN with FWHM$_{\rm H\beta}
\ltsim 2000\kms$ are classified as the narrow-line Seyfert~1 galaxies
(NLS1; Osterbrock \& Pogge 1985), and are distinguished from the bulk
of the Seyfert~1 galaxies.  In addition to the strong soft X-ray
excess below $2\kev$ and narrower permitted lines, NLS1 also show a
number of extreme properties e.g., steep $2-10\kev$ power-law
continuum, extreme X-ray variability and strong Fe~II emission.  These
properties are all related and one suggestion is that NLS1 are the
supermassive black hole analogue of stellar mass black hole X-ray
binaries (BHB) in their `high/soft' state.

However, the origin of the soft X-ray excess emission has remained a
major problem in AGN research over the past two decades.  The soft
excess emission is a smooth continuum component rather than a blend of
emission/absorption features as revealed by \chandra{} and \xmm{}
grating observations (e.g., Turner et al. 2001a; Collinge et al 2001).
When described as a thermal component, the soft excess emission has a
remarkably constant temperature ($\sim 100-200\ev$) across AGN with a
wide range of black hole masses (Czerny et al. 2003; Gierlinski \&
Done 2004; Crummy et al. 2006).  There are a number of ideas that have
been proposed to explain the origin of the soft excess component.
Pounds, Done \& Osborne (1995) pointed out the similarity of the X-ray
spectrum of NLS1 and Cyg X-1 in its high state (HS) and suggested that
the soft excess could be the optically thick emission from an
accretion disk. However, the observed soft excess emission is too hot
for a disk around a supermassive black hole. One possibility is that
there is a cool, optically thick Comptonizing region in addition to a
hot ($\sim 100\kev$), optically thin ($\tau\sim1$) region producing
the high energy power-law emission. The cool region can Comptonize the
disk photons and smoothly connect the disk emission up to soft X-ray
energies (Magdziarz et al.  1998).  Gierlinski \& Done (2004) tested
this model using high quality \xmm{} spectra of 26 radio-quiet PG
quasars. Although the spectra are well described by the Comptonization
model, the temperature of the cool Comptonizing region is remarkably
constant despite a large range in the black hole masses.  They
proposed that the observed soft excess is an artifact of heavily
smeared, strong partially ionized absorption (see also Gierlinski \&
Done 2006; Schurch \& Done 2006). Another possibility is
that the emission is dominated by ionized Compton reflection. This can
happen if the disk is clumpy at high accretion rates, thus hiding the
hard X-ray source among many clumps. The ionized reflection can
explain the observed soft excess emission (Fabian et al. 2002). The
last two models, based on relativistically smeared absorption and
ionized reflection, relate the soft excess emission to atomic
processes and therefore provide a natural explanation for its constant
temperature for AGN with a wide range in their black hole masses.

The three models, cool Comptonization, ionized reflection and heavily
smeared absorption, all result in statistically equally good fits.
The fitting criteria are not sufficient to discriminate among them
(Sobolewska \& Done 2005).  Variability properties of the soft and
hard X-ray emission and the relationship between them may provide
strong constraints on the above models for the soft excess emission. A
$35{\rm~day}$ long \asca{} monitoring observation of the NLS1 galaxy
Akn~564 revealed a distinction in the variability of the soft excess
and power-law components on a time scale of weeks, with the soft
excess emission varying by a factor of $\sim6$ compared to a factor of
$\sim4$ in the power-law component (Turner et al. 2001b). Similarly, a
long $\sim 10{\rm~days}$ \asca{} observation of the NLS1 galaxy
IRAS~13224-3809 suggested that the soft excess component dominates the
observed variability on a time scale of $\sim~a~week$, but on shorter
time scales ($\sim 20000\s$), the power-law component appears to
dominate the variability (Dewangan et al. 2002). A $12{\rm~day}$ long
\asca{} observation of another NLS1 galaxy Ton~S180 suggested that the
soft excess and the power-law fluxes are not well correlated on time
scales of $\sim 1000\s$ (Romano et al. 2002).  These long \asca{}
observations suggest two separate continuum component unlike the
heavily smeared, ionized absorption model based on a single continuum
component.  Gallo et al.  (2004) found evidence for alternate leads
and lags between the $0.3-0.8\kev$ and $3-10\kev$ band emission from
the NLS1 galaxy IRAS~13224-3809.  Dasgupta \& Rao (2006) reported
energy dependent delay between the soft and hard X-ray emission in
Mrk~110. These observations suggest different emitting regions for the
soft excess and the power-law component from NLS1 galaxies. In this
paper, we use \xmm{} observations of two NLS1 galaxies Akn~564 and
Mrk~1044 to study the variability properties of the soft and hard
X-ray emission and investigate the origin of the soft excess emission.
In Section 2, we outline the \xmm{} observations of Akn~564 and
Mrk~1044. We describe the temporal analysis in Section 3 and spectral
analysis in Section 4 and discuss the results in Section 5.

\setlength{\voffset}{0mm}
\begin{figure*}
  \centering \includegraphics[width=8cm]{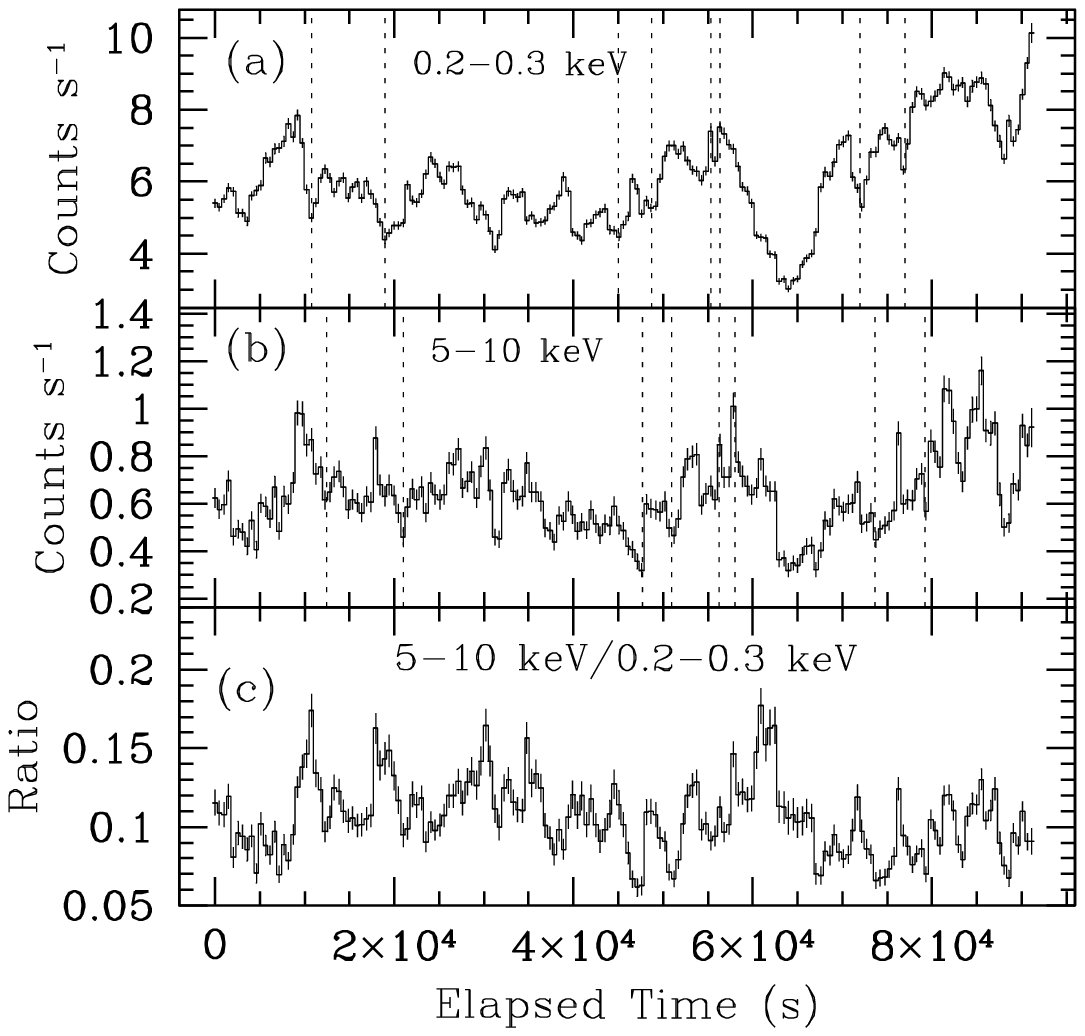}
  \includegraphics[width=8cm]{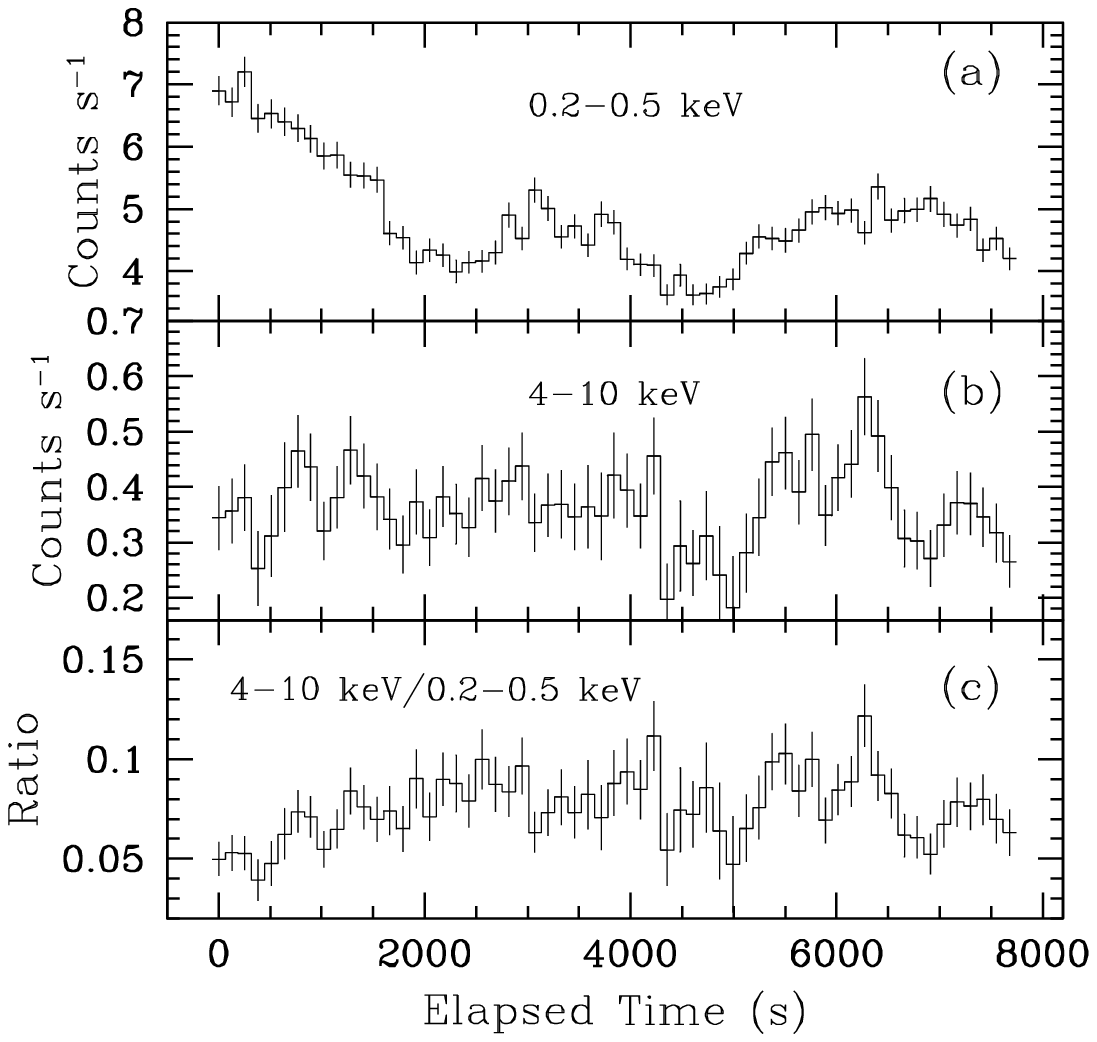}
  \caption{\xmm{} light curves and hardness ratio of Akn~564 and
    Mrk~1044. {\it Left:} Combined EPIC-pn and MOS light curves of
    Akn~564 in the bands $0.2-0.3\kev$ and $5-10\kev$ bands and the
    hardness ratio. The bin sizes are $512\s$. Vertical dotted lines
    mark the dips and peaks in the $0.2-0.3\kev$ and $5-10\kev$ light
    curves that show possible evidence for a delay.  {\it Right:}
    Combined EPIC-pn and MOS light curves of Mrk~1044 in the
    $0.2-0.5\kev$ and $2-10\kev$ bands and the hardness ratio. The bin
    sizes are $128\s$. The soft and hard band light curves do not
    appear to be strongly correlated. }
  \label{f1}
\end{figure*}
   
\section{Observation and Data Reduction}
Akn~564 was observed by \xmm{} on 2005 January 04 (ObsID: 0206400101)
for an exposure time of $100\ks$. Mrk~1044 was observed by \xmm{} on
2002 July 23 (ObsID: 0112600301) for $8\ks$. The EPIC-pn and MOS
cameras \citep{Struderetal01,Turneretal01c} were operated in the small
window mode using the medium filter during both the observations.  The
data were reduced using version 6.5.0 of the SAS software.
Examination of the background rate above $10\kev$ showed that the
observation of Akn~564 was partly affected by the flaring particle
background before an elapsed time of $7.5\ks$ and this early period
was therefore excluded to obtain a continuous exposure with steady
background.  For temporal analysis, we used all the pn and MOS events
with patterns $\le 12$ and a continuous exposure of $91.3\ks$ during
which both the pn and MOS cameras operated simultaneously.  There were
three short duration ($<1000\s$), small amplitude background flares
(maximum count rate $\sim 2{\rm~counts~s^{-1}}$ above $10\kev$) during
the observation of Mrk~1044. The varying background contribution to
the source was corrected by subtracting the background rate. For
spectral analysis, we chose a count rate cut-off criterion to exclude
the high particle background.  In case of Akn~564, the MOS data are
affected with photon pile-up. In order to achieve the best possible
spectral resolution and to have the best spectral calibration, we
chose the EPIC-pn data and
used the good X-ray events (FLAG=0) with pattern zero (single pixel
events) for Akn~564. For Mrk~1044, we used the good  EPIC-pn events with
pattern $0-4$ (single and double pixel events).  
We used circular regions with radii of $40\arcsec$,
centered at the peak position of Akn~564 and Mrk~1044, to extract the
source spectra.  We also extracted background spectra from appropriate
nearby circular regions, free of sources.

We also processed the RGS data obtained from the long observation of
Akn~564. We extracted the first order source and the background
spectra by making the spatial, order selections on the processed event
files, and calibrated by applying the most recent calibration data. We
also used a temporal filter to exclude the high particle background
based on a count rate cut-off criterion. We have used the ISIS
(version 1.4.7) spectral fitting environment(Houck \& Denicola 2000;
Houck 2002) for power and energy spectral analysis. Unless otherwise
stated, all errors are quoted at $90\%$ confidence level.

 \begin{figure}[H]
   \centering \includegraphics[width=7cm]{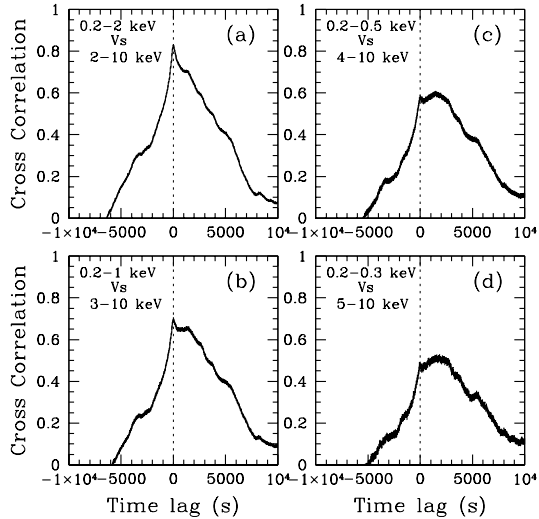}
   \caption{Cross correlation function between soft and hard X-ray
     light curves of Akn~564: (a) $0.2-2\kev - 2-10\kev$, (b)
     $0.2-1\kev$ Vs $3-10\kev$, (c) $0.2-0.5$ Vs $4-10\kev$ and (d)
     $0.2-0.3\kev$ Vs $5-10\kev$. The time delays refer to the soft
     band. Positive peaks imply that the variations in the hard band
     lag that in the soft band. }
   \label{f2}
 \end{figure}

\section{Temporal analysis}
\subsection{Akn~564}
For temporal analysis, we combined the pn and MOS data to increase the
signal-to-noise ratio. In order to study cross-correlation between
different energy bands, we extracted light curves in several energy
bands $0.2-0.3$, $0.2-0.5$, $0.2-1$, $0.2-2$, $2-10$, $3-10$, $4-10$
and $5-10\kev$ with time bins of $64\s$.  Figure~\ref{f1} shows the
light curves of Akn~564 in the $0.2-0.3$ and $5-10\kev$ bands rebinned
to $512\s$. X-ray emission from Akn~564 is highly variable.  The
$0.2-0.3\kev$ band light curve shows a trough-to-peak variations by a
factor of $\sim3.3$, while the hard band ($5-10\kev$) shows a
trough-to-peak variation by a factor of $\sim 3.6$. Although the
$0.2-0.3\kev$ and $5-10\kev$ band light curves appear to be
correlated, the most rapid variability is seen in the hard band. A
visual comparison of the light curves suggests that variations in the
hard band appear to lag those in the soft band. To investigate
further, we have performed cross-correlation and power spectral
analysis.

\subsubsection{Cross Correlation Analysis}
We have computed cross-correlation function (CCF) using the XRONOS
program {\tt crosscor}. We used the slow direct Fourier algorithm to
compute the CCF between the soft and hard band light curves with
$64\s$ bins.  Figure~\ref{f2} shows the CCFs as a function of time
delay between the energy bands (a) $0.2-2\kev$ and $2-10\kev$, (b)
$0.2-1\kev$ and $3-10\kev$, (c) $0.2-0.5\kev$ and $4-10\kev$ and (d)
$0.2-0.3\kev$ and $5-10\kev$. The $0.2-2\kev$ and $2-10\kev$ band
light curves are strongly correlated without any time delay as
indicated by the sharp peak at zero lag. However, the strength of the
sharp peak decreases with the increase in the separation between the
soft and hard bands. In the CCF of $0.2-1\kev$ and $3-10\kev$ band
light curves, the sharp peak weakens and another broad peak at
positive time lags becomes relatively stronger. The broad peak
dominates the CCF for the $0.2-0.5\kev$ and $4-10\kev$ band light
curves and only a weaker sharp peak is seen. The sharp peak is not
clearly seen in the CCF between $0.2-0.3\kev$ and $5-10\kev$ bands,
only the broad peak is seen. Thus the hard $4-10\kev$ or $5-10\kev$
band clearly lags the soft $0.2-0.5\kev$ or $0.2-0.3$ band.

\begin{figure*}
  \centering
  \includegraphics[width=6cm]{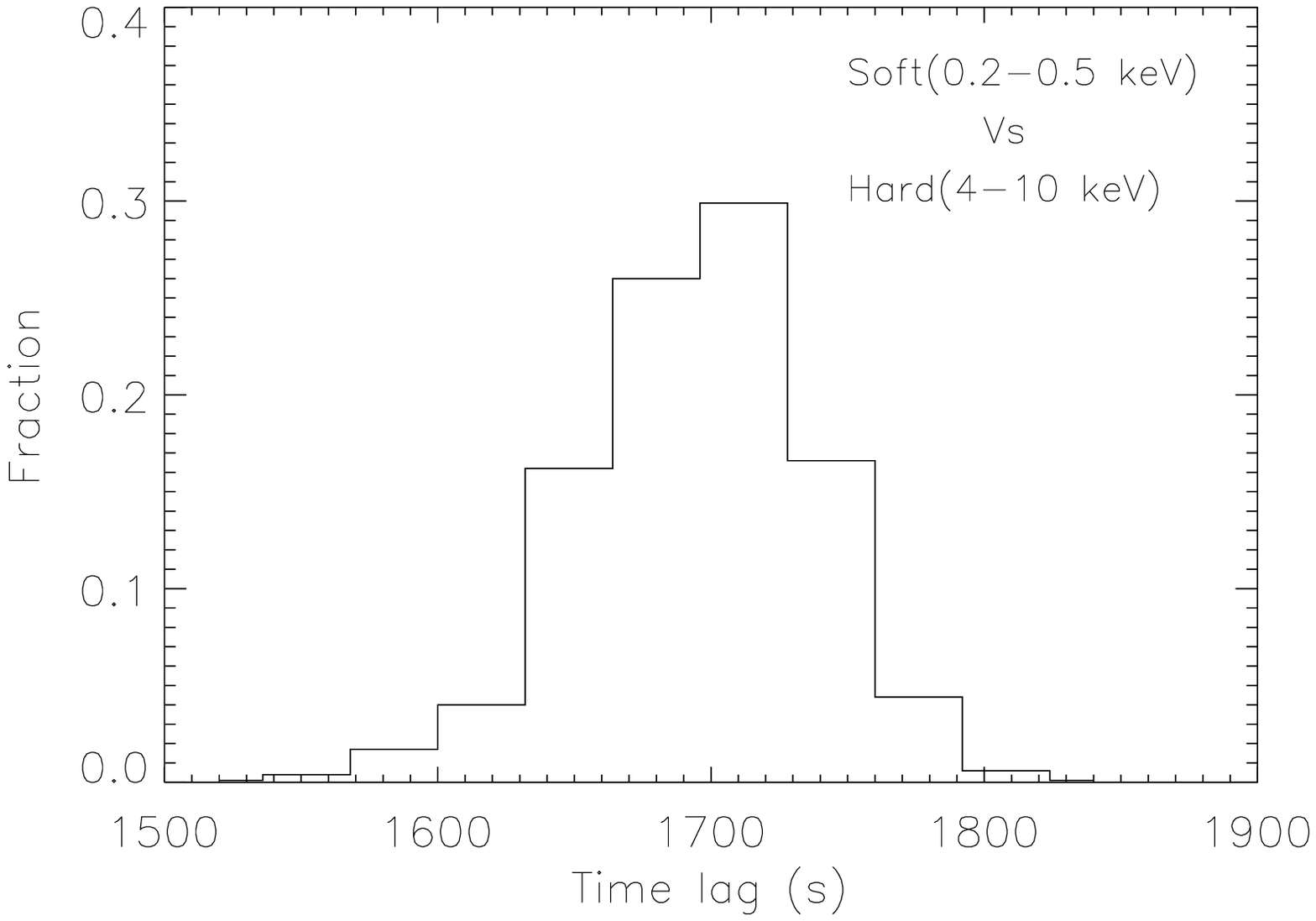}
  \includegraphics[width=6cm]{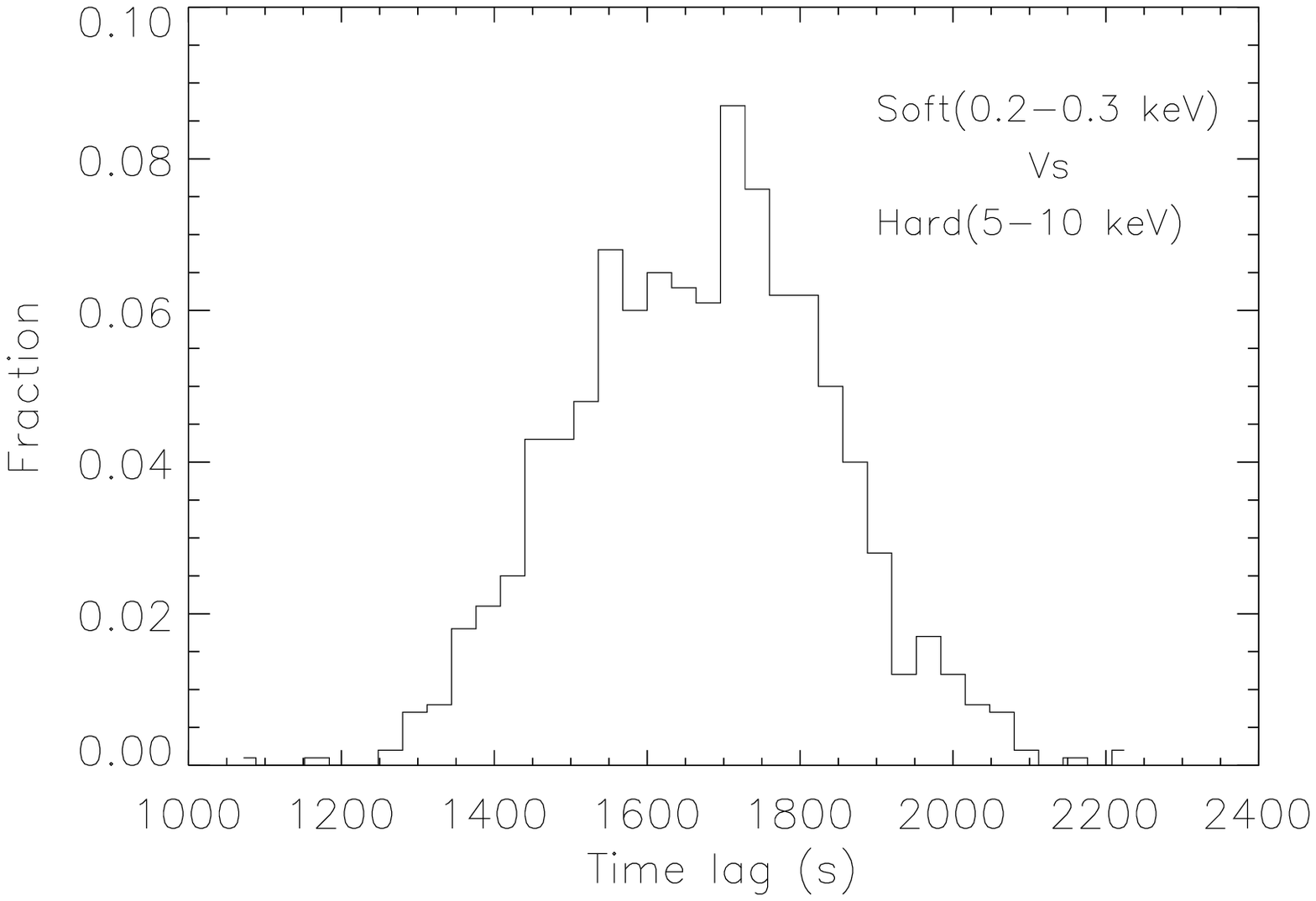}
  \caption{ Distributions of time lags obtained from the simulated
    soft and hard band lightcurves. {\it Left:} Time-lag distribution
    between $0.2-0.5$ and $4-10\kev$ bands. {\it Right:} Time-lag
    distribution between $0.2-0.3$ and $5-10\kev$ bands. }
  \label{f3}
\end{figure*}

It is hard to assess the statistical significance of the time lag
based on model fitting to the CCF using the traditional $\chi^2$
minimization technique. The reasons are ($i$) the shape of the CCF is
generally complex, ($ii$) the CCF data points are not independent and
($iii$) the errors are not normally distributed.  Therefore, we have
adopted a model independent way, following Peterson et al. (1998), to
measure the time lag. The principal source of uncertainty in the time
lag between two long, continuous light curves is the uncertainty in
the observed count rates. To account for this uncertainty, we have
performed Monte Carlo simulations using IDL (version 6.3). We
simulated 1000 pairs of soft and hard band lightcurves by modifying
the observed count rates by random Gaussian deviates assuming that the
errors on the observed count rates are normally distributed. We
calculated the time lag between each pair of the simulated soft and
hard band light curves as the centroid of the CCF based on all points
with correlation coefficients in excess of 0.4.  Figure~\ref{f3} shows
the distribution of time lags between the $0.2-0.5\kev$ and $4-10\kev$
bands. We calculated the time lag between the observed soft and hard
bands as the centroid of the CCF and the errors on the time lag by
directly integrating the time-lag distribution over a range to obtain
$68\%$, $90\%$ or $99\%$ confidence level.  Thus we measured the time
lag between the $0.2-0.5$ and $4-10\kev$ bands to be $1768\pm92\s$
($68\%$), $\pm122\s$ ($90\%$) or $\pm191\s$ ($99\%$).  A similar
procedure resulted in the time lag between the $0.2-0.3$ and
$5-10\kev$ to be $1767\pm196\s$ ($68\%$), $\pm320\s$ ($90\%$) or
$\pm462\s$ ($99\%$).

\begin{figure}[H]
   \centering
   \includegraphics[width=7cm]{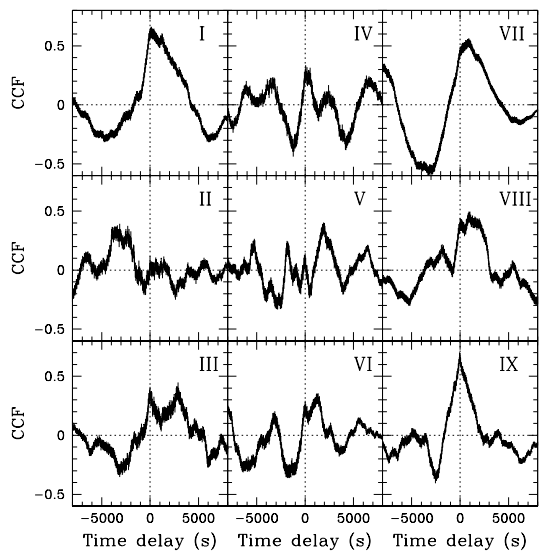}
   \caption{Cross correlation functions of $0.2-0.5\kev$ and
     $4-10\kev$ band lightcurves for Akn~564. The CCFs have been
     derived by splitting both $0.2-0.5\kev$ and $4-10\kev$
     lightcurves into nine segments. Each segment has a length of
     $10.1\ks$.}
   \label{f4}
 \end{figure}

We have also searched for variations in the CCF or time delay. For
this purpose, we divided both $0.2-0.5\kev$ and $4-10\kev$ light
curves into nine equal segments with lengths $10.1\ks$ and computed
the CCFs for each segment. The nine CCFs are plotted in
Figure~\ref{f4}. The CCFs are not similar. There is no or weak
correlation between the soft and hard band light curves in the
segments II-VI and VIII.  In segment IV, the soft and hard band appear
to be weakly anti-correlated with a soft band lead of $\sim 1200\s$.
The two bands are strongly correlated without any lag for the segments
I \& IX, while for the segments VII and VIII, the hard band lags
behind the soft band.  For these segments, the time lag varies from
$\sim 900-2900\s$. In segment VII, the soft and hard band appear to
anti-correlate at a lag of $\sim -3000\s$.  Brinkmann, Papadakis \&
Raeth (2007) have employed the sliding window technique to study the
cross-correlation properties of Akn~564 between the $0.3-1\kev$ and
$3-10\kev$ bands using the same data. There results are similar to the
presented here if we note that the first $7.5\ks$ of data were
excluded from our analysis.  The leads and lags in the X-ray emission
from Akn~564 are also similar to the alternate leads and lags in
$4500\s$ intervals found between the $0.3-0.8\kev$ and $3-10\kev$ band
X-ray emission from IRAS~13224-3809 by Gallo et al. (2004).

\begin{figure*}
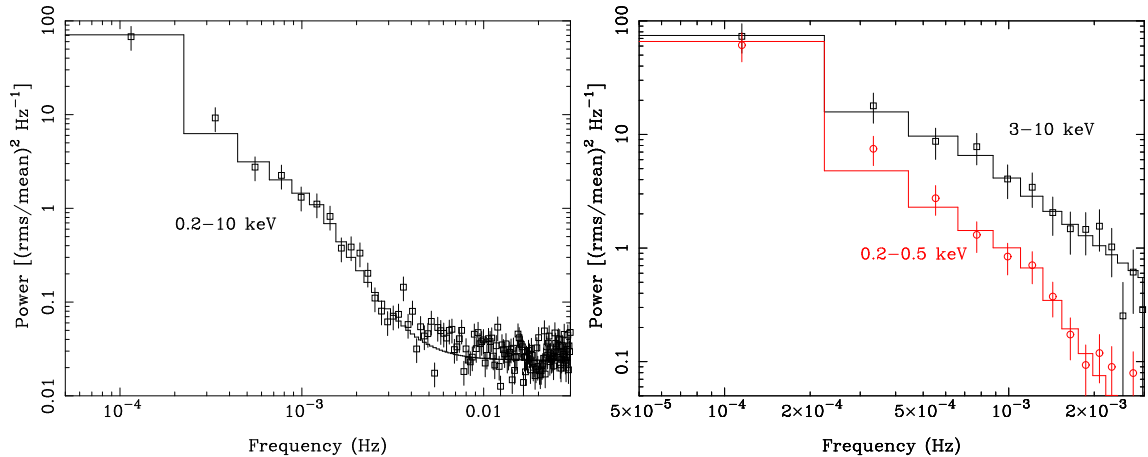

  \centering \includegraphics[width=6cm,angle=-90]{f5a.ps} \centering
  \includegraphics[width=6cm,angle=-90]{f5b.ps}
  \caption{{\it Left:} Power density spectra of Akn~564 in the full
    band $0.2-10\kev$. The contribution due to the Poisson noise has
    not been subtracted. {\it Right:} Power density spectra in the
    soft ($0.2-0.5\kev$) and hard ($3-10\kev$) bands, derived from the
    combined EPIC-pn and MOS data. The constant power expected from
    the Poisson errors have been subtracted. The best-fitting broken
    power-law models are also shown. The soft band power density
    spectrum is clearly steeper and weaker than the hard band power
    density spectrum. }
  \label{f5}
\end{figure*}

\subsubsection{The power density spectra}
To make a further study of the variability properties of Akn~564, we
have calculated power density spectra (PDS) of Akn~564 from the
combined pn+MOS lightcurves in the full ($0.2-10\kev$), soft
($0.2-0.5\kev$) and hard ($3-10\kev$) bands, sampled at $16\s$.  We
used the slow algorithm implemented in the XRONOS program `powspec'.
Following Papadakis \& Lawrence (1993), we rebinned the PDS in
logarithmic space with a binsize of 20 and performed the fitting
within the ISIS (version 1.4.5) spectral fitting environment. We
converted the PDS to equivalent energy spectra (power $\rightarrow$
counts/bin; $\hz \rightarrow \kev$) using the {\tt sitar} timing
package.  We fitted the PDS with a simple power-law model. We also
used a constant to account for the power arising from the Poisson
noise. The simple power-law model resulted in an unacceptable fit
(minimum $\chi^2 = 225.2$ for $139$ degrees of freedom (dof)). We then
replaced the power-law with a broken power-law model. This model
resulted in the minimum $\chi^2 = 169.2$ for $137$ dof.  The broken
power-law model provided an improved fit at a statistical significance
level of $>99.99\%$ compared to the power-law model according to the
maximum likelihood ratio test. Thus we conclude that the PDS of
Akn~564 has a break at $(1.2\pm0.3) \times 10^{-3}\hz$ with power-law
indices $1.3_{-0.2}^{+0.1}$ below the break and $3.2_{-0.4}^{+0.5}$
above the break.  The break frequency reported here is similar to the
high frequency break at $\sim 2\times 10^{-3}\hz$ discovered by
Papadakis et al. (2002) based on ASCA data (see also Alevaro et al.
2006) The constant power due to Poisson noise is
$0.024\pm0.0012{\rm~(rms/mean)^2 Hz^{-1}}$.
The PDS and the best-fit broken power-law plus constant model are
plotted in Figure~\ref{f5}. The power arising from the Poisson errors
have not been subtracted.

We also fitted the power-law and broken power-law models to the soft
($0.2-0.5\kev$) and hard ($3-10\kev$) PDS of Akn~564. The power-law
model resulted in an unacceptable fit to the PDS in the soft band
($\chi^2 = 193.2$ for $139$ dof) with power-law index of $1.7\pm0.1$.
The broken power-law model provided a better fit ($\chi^2/dof
=160.9/137$) with power-law indices $1.4_{-0.2}^{+0.1}$ and
$4.0_{-1.3}^{+1.8}$ before and after the break frequency
$1.1_{-0.6}^{+0.2}\times 10^{-3}\hz$. This is an improvement over the
power-law fit at a significance level of $>99.99\%$ based on the
maximum likelihood ratio test.  The power-law model provided an
acceptable fit to the hard band PDS ($\chi^2 = 138.4$ for $139$ dof)
with an index of $1.2\pm0.1$. The broken power-law model resulted in
an improved fit ($\chi^2/dof = 127.2/137$) with indices $<1.0$ ($90\%$
upper limit) and $1.8_{-0.4}^{+0.5}$ before and after the break
frequency $< 1.2\times 10^{-3}\hz$ ($90\%$ upper limit). This is an
improvement over the power-law model at a level of $99.6\%$
significance level. However, the parameters are not well determined.
We have plotted the soft and hard band PDS and the best-fit broken
power-law models in Fig.~\ref{f5} ({\it right panel}).  Clearly the
soft band PDS is steeper than the hard band PDS. This implies that the
X-ray emission in the hard band varies more than that in the soft band
on shorter time scales.

 \begin{figure}[H]
   \centering \includegraphics[width=7cm]{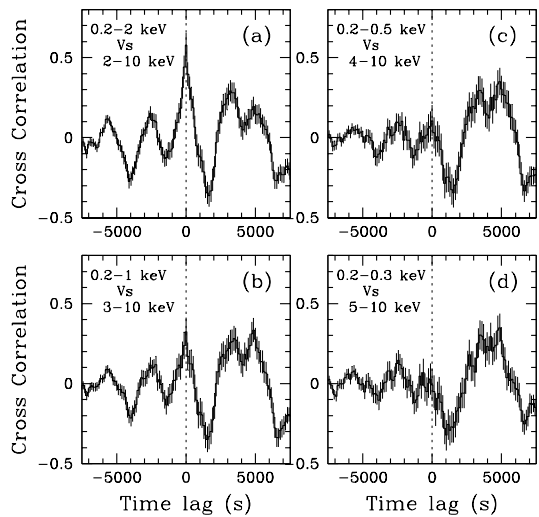}
   \caption{Cross correlation function between soft and hard X-ray
     light curves of Mrk~1044: (a) $0.2-2\kev - 2-10\kev$, (b)
     $0.2-1\kev$ Vs $3-10\kev$, (c) $0.2-0.5$ Vs $4-10\kev$ and (d)
     $0.2-0.3\kev$ Vs $5-10\kev$. The time delays refer to the soft
     band. Positive peaks imply that the variations in the hard band
     lag that in the soft band. }
   \label{f6}
 \end{figure}

\subsection{Mrk~1044}
As stated above, the observation of Mrk~1044 was affected by
short-duration, weak flares. Therefore, the source light curves were
corrected for background contribution using appropriate background
light curves extracted from blank sky regions. The combined pn+MOS and
background corrected light curves of Mrk~1044 are shown in
Fig.~\ref{f1} (right). X-ray emission from Mrk~1044 is variable with
trough-to-peak variations of $\sim 2$ ($0.2-0.5\kev$) and $\sim 3$
($4-10\kev$). However, the $0.2-0.5\kev$ and $4-10\kev$ light curves
do not appear to be strongly correlated.

The cross correlations between the soft and hard bands are plotted in
Figure~\ref{f6}. Similar to Akn~564, the $0.2-2\kev$ and $2-10\kev$
bands are strongly correlated without any delay and the strength of
this correlation decreases with the interval between the soft and hard
bands. There is no correlation between the $0.2-0.3\kev$ and
$5-10\kev$ bands at zero lag. There is a weak anti-correlation at a
positive delay of $\sim1000\s$ and a weak correlation at positive
delays of $3000-5000\s$. As the CCF for Mrk~1044 is based on a short
$7.9\ks$ observation, the CCF is likely only one representation of
many possible cross correlations similar to that seen for Akn~564.

\section{Spectral Analysis}
We have performed spectral modeling of Akn~564 and Mrk~1044. We
present the results based on the high signal-to-noise pn data for both
Akn~564 and Mrk~1044 and the high resolution RGS data for Akn~564. We
note that Papadakis et al.  (2007) have performed a detailed spectral
analysis of Akn~564 using the same observation. The purpose here is to
investigate the alternative models for the origin of the soft X-ray
excess emission.  All the spectra were analyzed with the Interactive
Spectral Interpretation System ({\tt ISIS, version 1.4.5}).  The pn
data were grouped to a minimum of $100$ and $20$ counts per spectral
channel for Akn~564 and Mrk~1044, respectively. We consider a model
fit as statistically unacceptable if the the null hypothesis can be
rejected at more than $95\%$ significance.  The errors on the best-fit
spectral parameters are quoted at a $90\%$ confidence level. Below we
use narrow Gaussian line components with their widths
fixed at $\sigma = 10\ev$ that are unresolved by the EPIC-pn data.

\subsection{The soft X-ray excess emission}
\begin{figure}[H]
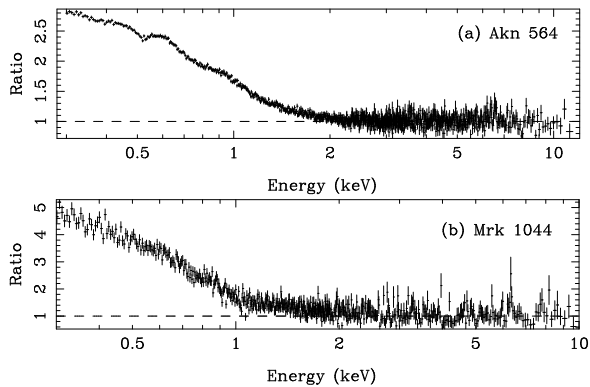

   \centering
   \includegraphics[width=2.5cm,angle=-90]{f7a.ps}
   \includegraphics[width=2.5cm,angle=-90]{f7b.ps}
   \caption{Ratios of observed EPIC-pn data for Akn~564 and Mrk~1044
     and the corresponding best-fit  power-law model fitted above
     $3\kev$ and
     extrapolated to lower energies.}
   \label{f7}
 \end{figure}

\begin{figure*}
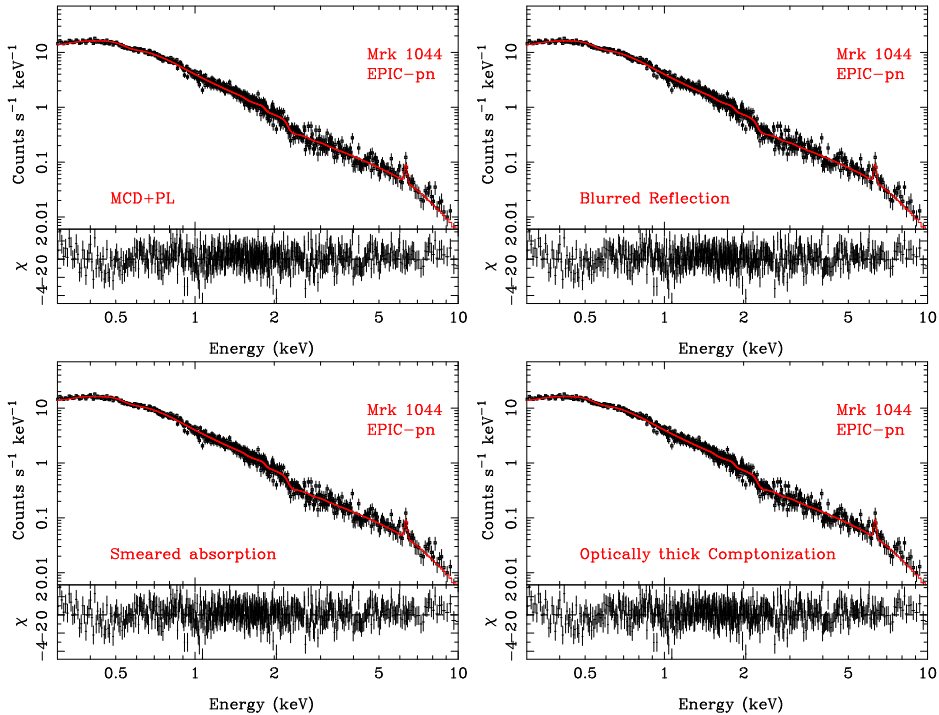

  \centering
  \includegraphics[width=4.7cm,angle=-90]{f9a.ps}
  \includegraphics[width=4.7cm,angle=-90]{f9b.ps}
  \includegraphics[width=4.7cm,angle=-90]{f9c.ps}
  \includegraphics[width=4.7cm,angle=-90]{f9d.ps}
  \caption{Observed EPIC-pn data and the best-fit models for Mrk~1044
    corresponding to the best-fit models and their parameters listed
    in Table~\ref{t1}.}
  \label{f9}
\end{figure*}

NLS1 galaxies are known to show strong soft X-ray excess emission
below $\sim 2\kev$. To show this soft excess, first we fitted a simple
absorbed power law (PL) model to the EPIC-pn spectra above $3\kev$
($3-10\kev$ for Mrk~1044 and $3-11\kev$ for Akn~564).  The simple PL
model resulted in minimum $\chi^2 = 527.4$ for $519$ dof and $119.0$
for $127$ dof for Akn~564 and Mrk~1044, respectively, thus providing
good fits to the $3-10\kev$ pn data.  We have plotted the ratio of the
pn data and the best-fit PL model in Figure~\ref{f7} after
extrapolating the model to low energies. The plots clearly show strong
soft excess emission below $\sim 2.5\kev$ and possible iron K$\alpha$
lines near $\sim 6.4\kev$ in the spectra of both the NLS1. The soft
X-ray excess emission from NLS1 are usually well described by a simple
blackbody or a multicolor disk blackbody (MCD). Addition of the MCD
component to the absorbed PL improved the fit. The fit was reasonably
good for Mrk~1044 ($\chi^2/dof = 628.0/541$). Addition of a narrow
Gaussian line at $\sim 6.4\kev$ further improved the fit ($\chi^2/dof
= 611.7/539$). The best-fit parameters of the MCD+PL model for
Mrk~1044 are listed in Table~\ref{t1} and the EPIC-pn data, the
best-fit model and residuals are plotted in Figure~\ref{f9}.

The MCD+PL model was a poor fit for Akn~564 ($\chi^2/dof =
1494.1/913$).  Examination of the residuals showed broad
emission/absorption features below $\sim 1\kev$ in the spectrum of
Akn~564. To verify and investigate these features, we have analyzed
the high resolution soft X-ray spectra of Akn~564 obtained with the
two RGS simultaneously with the EPIC-pn data. We used the Cash
statistic to fit the low count RGS data.  We fitted a simple PL model,
modified with the neutral absorption, jointly to RGS1 and RGS2 data in
the $0.35-1.9\kev$ band.  This resulted in a steep spectrum ($\Gamma
\sim 3.0$) and $N_H = 7.6\times10^{20}{\rm~cm^{-2}}$. Addition of an
MCD component improved the fit ($C/dof=2575.9/1572$; $\Delta C =
-225.4$ for two parameters).  The best-fit parameters are $N_H \sim
5.2\times 10^{20}{\rm~cm^{-2}}$, $\Gamma \sim 2.44$ and the MCD
$kT_{in}\sim 165\ev$.  Examination of the ratio of the data and model
revealed a strong absorption feature at $\sim 0.5\kev$ and additional
weak and narrow absorption features in the $0.5-1\kev$ band,
suggesting the presence of a warm absorber medium in Akn~564.

\begin{figure*}
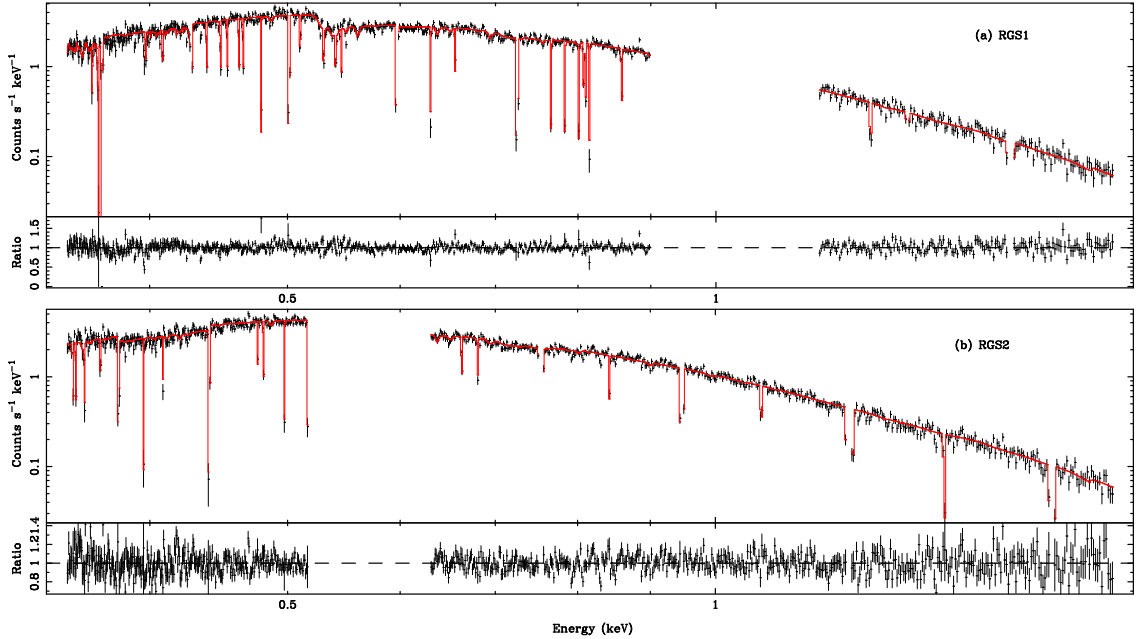

  \centering
  \includegraphics[height=15cm,angle=-90]{f8a.ps}
  \includegraphics[height=15cm,angle=-90]{f8b.ps}
  \caption{(ab) RGS1 data, the best-fitting joint model and the ratios
    of the observed data and the model for Akn~564. (b) The same as
    (a) but for RGS2. The best-fit model was derived by joint spectral
    fitting to both RGS1 and RGS2 data.}
  \label{f8}
\end{figure*}

\begin{table*} 
\centering 
{\small \caption{Spectral model
      parameters for Akn~564 and Mrk~1044.  \label{t1}}
    \begin{tabular}{lcc} \tableline\tableline Parameter &
      Akn~564\tablenotemark{1} & Mrk~1044 \\ \tableline & MCD$+$PL &
      \\ \tableline
      $N_H$ ($10^{20}{\rm~cm^{-2}}$) & $5.24$(f)   & $3.54$(f) \\
      $kT_{in}$ (eV)                & $179\pm2$  & $138\pm3$  \\
      $n_{MCD}$\tablenotemark{2}    & $3002_{-100}^{+104}$   & $2992_{-277}^{+316}$  \\
      $\Gamma$                     & $2.48\pm0.01$  &  $2.14\pm0.04$ \\
      $n_{PL}$\tablenotemark{3}    & $(1.35\pm0.02)\times10^{-2}$ & $(3.2\pm0.1)\times10^{-3}$  \\
      $E_{FeK\alpha}$ (keV)             &   $6.72\pm0.03$  & $6.42\pm0.04$ \\
      $f_{FeK\alpha}$\tablenotemark{4}  & $(5.1\pm1.8)\times10^{-6}$ & $(1.14\pm0.46)\times 10^{-5}$ \\
      $f_{0.3-10keV}$\tablenotemark{5} & $6.6\times10^{-11}$ & $2.1\times10^{-11}$  \\
      $\chi^2/dof$ & $919.3/854$ & $611.7/539$ \\ \tableline & Blurred
      Reflection \\ \tableline
      $N_H$ ($10^{20}{\rm~cm^{-2}}$)  & $5.24$(f)   & $3.54$(f) \\
      $\Gamma$   & $2.468_{-0.001}^{+0.002}$ & $2.20_{-0.004}^{+0.002}$  \\
      $n_{PL}$\tablenotemark{3}   & $9.29\pm0.03\times10^{-3}$ & $1.43_{-0.03}^{+0.05}\times10^{-3}$  \\
      $\beta$      & $\ge10.0$   & $\ge10.0$  \\
      $r_{in}$($r_g$)   &  $<1.58$   & $<1.25$    \\
      $i$       &   $66\pm1^\circ$        &  ${65.7_{-0.2}^{+0.3}}^\circ$   \\
      $n_{refl}$\tablenotemark{6} & $(4.27\pm0.02)\times10^{-7}$  & $1.7_{-0.01}^{+0.02}\times10^{-6}$  \\
      $A_{Fe}$                   & $0.69\pm0.01$ &  $0.35_{-0.03}^{+0.04}$  \\
      $\xi {\rm~(erg~cm~s^{-1})}$                     & $3452_{-114}^{+129}$  &  $334.5_{-1.2}^{+2.1}$  \\
      $E_{FeK\alpha}$ (keV) & $6.72_{0.03}^{+0.05}$ &  $6.42\pm0.04$ \\
      $f_{FeK\alpha}$\tablenotemark{4}  &  $2.5_{-1.5}^{+2.1}\times10^{-6}$ & $1.0_{-0.5}^{+0.4}\times10^{-5}$  \\
      $f_{0.3-10keV}$\tablenotemark{5}  & $6.4\times10^{-11}$ & $2.0\times10^{-11}$  \\
      $\chi^2/dof$ & $954.2/850$ & $671.6/535$ \\ \tableline & Smeared
      absorption \\ \tableline
      $N_H$ ($10^{20}{\rm~cm^{-2}}$)  & $5.24$(f)   & $3.54$ (f) \\
      $\Gamma$   & $2.783_{-0.008}^{+0.006}$ &  $2.49\pm0.02$ \\
      $n_{PL}$\tablenotemark{3}   & $(3.67\pm0.05)\times10^{-2}$  & $(7.6\pm0.2)\times10^{-3}$   \\
      wind $N_H$ ($10^{22}{\rm~cm^{-2}}$) & $19.6_{-0.9}^{+1.8}$ & $15.7_{-1.6}^{+1.8}$ \\
      $log (\xi/{\rm erg~cm~s^{-1}})$  & $2.9_{-0.03}^{+0.05}$  & $2.89\pm0.04$ \\
      $\sigma$\tablenotemark{7}   & $0.81\pm0.03$  &  $0.29\pm0.03$ \\
      $E_{FeK\alpha}$ (keV) &$6.72\pm0.03$  &  $6.42\pm0.04$ \\
      $f_{FeK\alpha}$\tablenotemark{4}  & $(5.1\pm2.6) \times 10^{-6}$  &  $(1.2\pm0.5)\times10^{-5}$ \\
      $f_{0.3-10keV}$\tablenotemark{5}  & $7.1\times10^{-11}$  & $2.1\times10^{-11}$  \\
      $\chi^2/dof$ & $1147.1/853$ & $592.4/538$ \\ \tableline &
      Thermal Comptonization \\ \tableline
      $N_H$ ($10^{20}{\rm~cm^{-2}}$)  & $5.24(f)$   & $3.54$ (f) \\
      $\Gamma$   & $2.46\pm0.01$ &  $2.11\pm0.05$ \\
      $n_{PL}$\tablenotemark{3}  & $(1.31\pm0.02)\times10^{-2}$  & $(3.1\pm0.15)\times10^{-3}$  \\
      $kT_{seed}$($\ev$) &  $30$(f)  & $30$(f)  \\
      $kT_e$($keV$)    & $0.176\pm0.006$   &  $0.137_{-0.010}^{+0.012}$  \\
      $\Gamma_{thcomp}$     & $2.1\pm0.1$  & $2.1\pm0.2$  \\
$E_{FeK\alpha}$ (keV) & $6.72\pm0.03$  &  $6.42\pm0.04$ \\
$f_{FeK\alpha}$\tablenotemark{4}  & $(4.9\pm1.8)\times10^{-6}$ & $(1.1\pm0.5)\times10^{-5}$  \\
 $f_{0.3-10keV}$\tablenotemark{5}  & $6.7\times10^{-11}$  &  $2.1\times10^{-11}$  \\
$\chi^2/dof$  & $900.2/853$  & $603.0/538$  \\ \tableline
\end{tabular}
\tablenotetext{1}{For
  Akn~564, all models were modified by the two phase warm absorber medium
  inferred from the RGS data and the EPIC-pn data were  fitted in the
  $0.6-11\kev$ band.}
\tablenotetext{2}{MCD normalization
  $n_{MCD}=(R_{in}/km)/(D/10{\rm~kpc})$, where $R_{in}$ is the inner
  radius and $D$ is the distance.}
\tablenotetext{3}{Power-law normalization in units of {$\rm photons~keV^{-1}~cm^{-2}~s^{-1}$} at $1\kev$.}
\tablenotetext{4}{Line flux in $\rm photons~keV^{-1}~cm^{-2}~s^{-1}$.}
\tablenotetext{5}{Observed flux in units of ${\rm~ergs~cm^{-2}~s^{-1}}$
  in the $0.3-10\kev$ band.}
\tablenotetext{6}{Normalization of the reflected spectrum.}
\tablenotetext{7}{Gaussian sigma for velocity smearing in units of $v/c$.}
}
\end{table*}

\subsection{Warm absorber model for Akn~564}
To describe the weak absorption features seen in the RGS data, we have
created a warm absorber model for Akn~564 using the spectral
simulation code CLOUDY version 7.02.01 (last described by Ferland et
al. 1998).  We used a multicomponent ionizing continuum similar to
that observed from Akn~564. We used the `agn' continuum available
within CLOUDY. This continuum has four parameters: the temperature of
the big blue bump emission, the slope of the high energy X-ray
continuum, X-ray to UV ratio ($\alpha_{ox}$) and the UV spectral
index. We set the temperature of the big blue bump component and the
slope of the high energy continuum to be similar to the temperature of
the soft X-ray excess emission and the slope of the hard power law,
respectively, derived from the EPIC-pn data.  For Akn~564,
$\alpha_{ox} = 0.941$ (Gallo 2006) and we used a typical value
of $-0.5$ for the UV spectral index (Elvis et al. 1994; Francis 1993).
The X-ray power law is assumed to fall as $\nu^3$ above $100\kev$ and
the big blue bump component is assumed to have an infra-red
exponential cut-off at $kT_{IR} = 0.01{\rm~Ryd}$ or $0.136\ev$. We
also included the cosmic microwave background radiation so that the
incident continuum has nonzero intensity at very long wavelengths. We
assumed a plane parallel geometry and calculated grids of models by
varying the ionization parameter and the total hydrogen column
density. We also included UTA features from Gu et al. (2006) in our
calculation. The grid of models were imported to ISIS in the form of
an XSPEC-style multiplicative table model as described in Porter et
al.  (2006).

To describe the weak absorption features in the RGS spectra of
Akn~564, we multiplied the CLOUDY warm absorber model to the MCD+PL
model.  The model was also modified by neutral absorption along the
line of sight as before. The addition of the warm absorber component
improved the fit ($C/dof=2469.3/1569$; $\Delta C = -106.6$ for three
additional parameters) with warm absorber column, $N_W \sim 4\times
10^{20}{\rm~cm^{-2}}$, ionization parameter ($log\xi \sim 2$) and the
outflow velocity ($v \sim 300\kms$). An additional warm absorber
component further improved the fit ($\Delta C = -70.5$ for three
parameters) with $N_W \sim 2\times 10^{-20}{\rm~cm^{-2}}$, $log \xi <
0.3$ and $v \sim -1000\kms$. Examination of the residuals showed a
strong absorption feature near $0.5\kev$ at the location of the
neutral oxygen edge. However, the shape of the absorption feature does
not resemble to an absorption edge, it is similar to an absorption
line with a very weak emission line at the center. This feature is
likely due to possible RGS calibration errors at the oxygen edge
and/or from the Galactic oxygen along the line of sight. Addition of a
Gaussian absorption line improved the fit ($\Delta C = -97.3$ for
three parameters). The parameters of the best-fit model to the RGS
data is listed in Table~\ref{t2}. We note that the best-fit neutral
absorption column density derived from the RGS data is $N_H =
5.24_{-0.03}^{+0.13}\times10^{20}{\rm~cm^{-2}}$ which is less than the
Galactic column of $6.40\times10^{20}{\rm~cm^{-2}}$ derived from the
HI map of Dickey \& Lockman (1990). The best-fit column, however, is
consistent with the Galactic column of
$5.34\times10^{20}{\rm~cm^{-2}}$ derived from the
Leiden/Argentine/Bonn (LAB) Survey of Galactic HI (Kalbera et al.
2005). McKernan, Yaqoob, \& Reynolds (2007) derived cold $N_H =
4_{-1}^{+2}\times10^{20}{\rm~cm^{-2}}$ using the Chandra high energy
transmission grating observations of Akn~564. Given the above
uncertainties in the measurement of cold $N_H$, we fix the cold
absorption at the best-fit RGS value of
$N_H=5.24\times10^{20}{\rm~cm^{-2}}$ in all the subsequent fits for
Akn~564.

\begin{table*}
\centering
{\small  \caption{Best-fit model parameters derived from the high resolution
    RGS observation of Akn~564 \label{t2}}
  \begin{tabular}{lll} \tableline\tableline
    Neutral absorption & $N_H$ ($10^{20}{\rm~cm^{-2}}$) &
    $5.25_{-0.03}^{+0.13}$  \\ \\
    Warm absorber I &  $N_W$ ($10^{20}{\rm~cm^{-2}}$) & $3.9_{-0.3}^{+0.5}$ \\
    & $log \xi$   &  $2.0\pm0.1$ \\
    & ($v/\kms$)~\tablenotemark{a}  & $-280^{+130}_{-180}$  \\ \\
    Warm absorber II &  $N_W$ ($10^{20}{\rm~cm^{-2}}$) & $2.1_{-0.1}^{+0.2}$ \\
    & $log \xi$   & $<0.32$  \\
    & ($v/\kms$)~\tablenotemark{a}  & $-1080^{+175}_{-155}$ \\ \\
    Disk blackbody  & $kT_{in}$ (eV)                & $171\pm2$  \\
    & $n_{MCD}$\tablenotemark{b}     & $2917_{-69}^{+18}$   \\ \\
    Power law  & $\Gamma$                     & $2.48\pm0.01$  \\
    & $n_{PL}$\tablenotemark{c}     & $(1.50\pm0.01)\times10^{-2}$
    \\ \\
    Absorption line & $E$ (keV)            &  $0.541\pm0.001$ \\
    & $f_{line}$\tablenotemark{d}  & $-2.6_{-0.3}^{+0.5}\times 10^{-4}$  \\
    $C/dof$     &              & $2301.5/1577$  \\ \tableline
\end{tabular}
\tablenotetext{a}{Velocity with respect to the systemic
  velocity. Negative sign indicates outflow.}
\tablenotetext{b}{MCD normalization
  $n_{MCD}=(R_{in}/km)/(D/10{\rm~kpc})$, where $R_{in}$ is the inner
  radius and $D$ is the distance.}
\tablenotetext{c}{Power-law normalization in units of {$\rm photons~keV^{-1}~cm^{-2}~s^{-1}$} at $1\kev$.}
\tablenotetext{d}{Line flux in $\rm photons~keV^{-1}~cm^{-2}~s^{-1}$.}
}
\end{table*}

The above analysis clearly shows the presence of a two phase warm
absorber medium in Akn~564. Some or all of the emission/absorption
features resulting in the poor quality of the fit to the EPIC-pn,
described above in section 4.1 are likely due to the presence of warm
absorbers. However, it is difficult to  determine accurately the
parameters of the warm absorber based on the lower resolution EPIC-pn
data alone.  Therefore, we used the best-fit warm absorber components
derived from the RGS data to model the EPIC-pn data. We used the two
component warm absorber model with the fixed parameters along with the
MCD+PL model and fitted the EPIC-pn data. The addition of the warm
absorber components improved the fit, providing $\chi^2/dof =
1654.1/912$.  However, the fit is still not acceptable
mainly due to a strong absorption feature near $0.5\kev$ and slight
excess of emission below $0.5\kev$. The absorption feature is stronger
by a factor of $\sim 2.5$ in the EPIC-pn data than that derived from
the RGS data.  Also the excess emission below $0.5\kev$ is not evident
in the RGS data. Thus EPIC-pn and RGS do not agree well below
$0.6\kev$. Therefore we ignore the EPIC-pn data below $0.6\kev$ in 
the fits for Akn~564 described below.

After excluding the data below $0.6\kev$, the MCD+PL model modified
with the two component warm absorber and neutral absorption along the
line of sight provided a much better fit ($\chi^2/dof =
941.2/856$). Adding a narrow Gaussian line at $\sim 6.6\kev$ further
improved the fit ($\Delta chi^2/dof = -21.9$ for two parameters).
Figure~\ref{f10} shows the EPIC-pn data, the best-fit model and the
$\chi$-residuals and the best-fit parameters are listed in
Table~\ref{t1}. The parameters of the MCD and PL components are
similar to that derived from the RGS data (see Table~\ref{t2} and
\ref{t2}).
The MCD component used above to describe the soft X-ray excess
emission is physically inappropriate.  The inner accretion disk
temperatures, $kT_{in} = 179\pm2\ev$ for Akn~564 and $138\pm3\ev$ for
Mrk~1044 are much higher than that expected from a disk around a black
hole of mass $10^6 - 10^7{\rm~M_{\odot}}$.  For this reason, the
origin of the soft excess emission has remained unsolved for more than
two decades.  

\subsection{Relativistically blurred ionized reflection}

\begin{figure*}
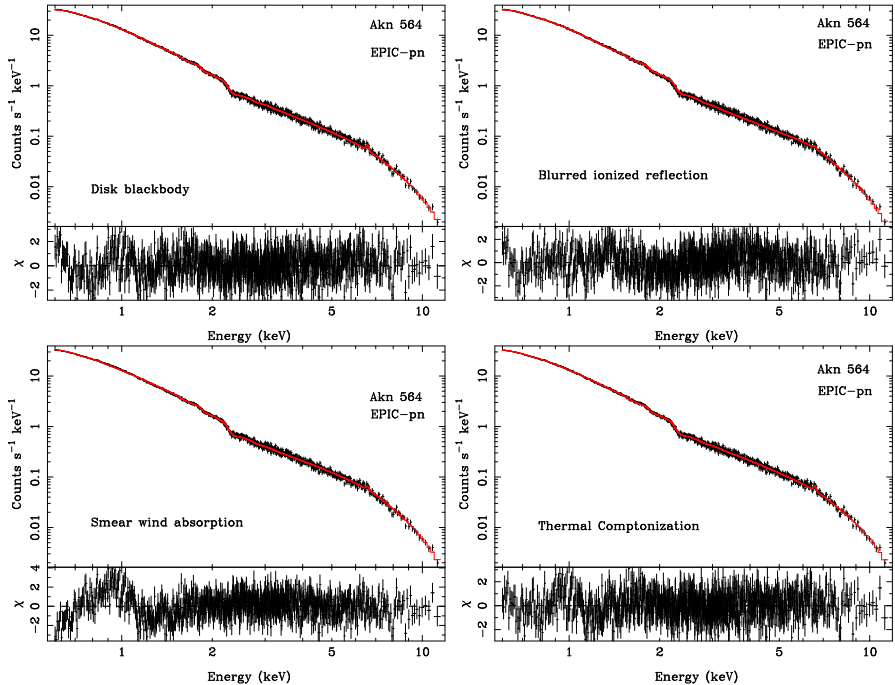

  \centering
  \includegraphics[width=4.5cm,angle=-90]{f10a.ps}
  \includegraphics[width=4.5cm,angle=-90]{f10b.ps}
  \includegraphics[width=4.5cm,angle=-90]{f10c.ps}
  \includegraphics[width=4.5cm,angle=-90]{f10d.ps}
  \caption{Observed EPIC-pn data and the best-fitting MCD+PL, blurred
    ionized reflection, smeared absorption from partially ionized wind
    and optically thick thermal Comptonization models for Akn~564. The
    best-fit parameters are listed in Table~\ref{t1}.}
  \label{f10}
\end{figure*}

One possibility is that the reflected emission from a partially
ionized accretion disk could be strong below $1\kev$ and can account
for the strong soft excess emission (see e.g., Crummy et al.  2005).
Though there is no clear evidence for a broad iron K$\alpha$ line,
such a line from an accretion disk could be hidden in the strong X-ray
continuum. To test whether the strong soft X-ray excess emission from
Akn~564 and Mrk~1044 could be produced by the ionized reflection
emission arising from the reprocessing of the primary X-ray emission
in the disk, we have fitted the relativistically blurred ionized
reflection model to the EPIC-pn spectra of both the NLS1. We used the
table model {\tt reflion} (Ross \& Fabian 2005; Ross, Fabian \& Young
1999) for the ionized reflection and blur the ionized reflection
emission relativistically by convolving with a {\tt LAOR} kernel ({\tt
  kdblur}) to obtain the relativistically broaden ionized reflection
from the inner regions of an accretion disk. The parameters of {\tt
  reflion} model are ionization parameter ($\xi$), iron abundance
relative to solar ($A_{Fe}$), photon index of the illuminating
power-law and the normalization of reflected spectrum.  The parameters
of the convolution model {\tt kdblur} are the inner and outer radii of
the disk ($r_{in}$ and $r_{out}$), emissivity index ($\beta$) and the
disk inclination ($i$).  We fixed the outer radius at $r_{out} =
400r_g$ where $r_g = GM/c^2$ is the gravitational radius. We used the
power-law component for the continuum.  We also used a narrow Gaussian
to describe the unresolved iron K$\alpha$ line from distant matter. We
also used the two component warm absorber model with the parameters
fixed at the values derived from the RGS data. We refer the full model
as the blurred reflection model. The best-fit parameters of this model
are listed in Table~\ref{t1}. The observed data and the best-fit
reflection models are shown in Figure~\ref{f9} (Mrk~1044) and
Figure~\ref{f10} (Akn~564). The model resulted in poorer fit than the
MCD+PL model for both the NLS1 ($\chi^2/dof = 954.2/837$ for Akn564
and $671.6/535$ for Mrk~1044).  Also the emissivity index is
unphysically large ($\beta \ge 10$) and the inner radius is very small
for both the AGN, suggesting that the model cannot account for the
observed smoothness of the soft excess X-ray emission.  The very steep
emissivity indices and the small inner radii of the LAOR kernel were
required to smooth the recombination features sufficiently to produce
the featureless soft X-ray excess emission from both the NLS1s.

\subsection{Smeared absorption from partially ionized material}
The observed soft excess emission could be an artifact of smeared
absorption from partially ionized material as described by Gierlin'ski
\& Done (2004). To test this scenorio, we have fitted the EPIC-pn data
with the smeared absorption model. We used the XSPEC local model {\tt
  swind1} for the smeared absorption from a partially ionized wind
from an accretion disk and the simple PL model for the continuum.
Again we used a Gaussian for the narrow iron K$\alpha$ line and the
two component warm absorber model with fixed parameters derived from
the RGS data. The parameters of the smeared wind model are absorption
column density, the ionization parameter ($\xi=L/nr^2$) and the
Gaussian sigma for velocity smearing in units of $v/c$. The smeared
absorption model provided a good fit to the spectrum of
Mrk~1044 ($\chi^2/dof = 592.4/538$) but fails to describe the spectrum
of Akn~564 ($\chi^2/dof = 1147.1/842$).  This model requires
unphysically large smearing velocity ($\sim 0.8c$) for Akn~564 to
smooth sufficiently the absorption lines in order to produce the
observed smooth soft excess emission (see Table~\ref{t1} and
Fig.~\ref{f10}).

\subsection{Optically-thick, thermal Comptonization}
The strong and smooth soft excess emission from NLS1 can also be
described as arising from thermal Comptonization in an optically thick
corona. To test such a scenario, we used  the thermal Comptonization model
(ThComp) described by Zdziarski et al (1996).  The free parameters of
ThComp model are the asymptotic power-law index ($\Gamma_{thcomp}$),
electron temperature ($kT_e$) and the seed photon temperature
($kT_{bb}$). The electron scattering optical depth ($\tau$) can be
calculated from the asymptotic power-law photon index
($\Gamma_{compth}$) and electron temperature ($kT_e$) as follows
\begin{equation}
  \alpha = \left[\frac{9}{4} + \frac{1}{(kT_e / m_e c^2) \tau (1 + \tau / 3)}\right]^{1/2} - \frac{3}{2}
\end{equation}
with $\Gamma_{thcomp} = \alpha + 1$ (Sunyaev \& Titarchuk 1980).  
We also used the
PL model to describe the hard PL for the hot Comptonized
component. We fixed the temperature of the seed photons for the
optically thick Comptonization at $30\ev$. This is equivalent to
assuming that the seed photons for the Comptonizations in the warm and
hot plasma arise from the accretion disk. Again we used the narrow
Gaussian line for the iron K$\alpha$ line. 
This model provided good fits for both
the NLS1s ($\chi^2/dof = 900.2/842$ for Akn~564 and
$\chi^2/dof=603.0/538$ for Mrk~1044).  The best-fit parameters are
listed in Table~\ref{t1} and the  EPIC-pn data and the best-fit
model are plotted in Fig.~\ref{f9} and Fig.~\ref{f10} .  The electron
temperatures are low ($\sim 0.18\kev$ for Akn~564 and $\sim0.14\kev$
for Mrk~1044) and electron scattering optical depths are very high
($\tau \sim 45$) for both the NLS1s.

We note that the thermal Comptonization model best describes the
EPIC-pn data in the $0.6-11\kev$ band for Akn~564 (see Table~\ref{t1}).
We expect that the most suitable model must provide a reasonably good
fit to the full band EPIC-pn data except for the absorption feature at
$0.54\kev$, provided that the calibration errors at low energies
($<0.5\kev$) are at a level of a few percent. We find that the thermal
Comptonization model best describes the full band data with
$\chi^2/dof = 1164.8/911$ without any significant change in the
best-fit parameters listed in Table~\ref{t1}. Addition of an
absorption line at $0.54\kev$ improved the fit to $\chi^2/dof =
985.5/908$ with line center at $E=0.54\pm0.005$, $f_{line} = -
(4.44\pm0.06)\times 10^{-4}{\rm~ergs~cm^{-2}~s^{-1}}$ and width
$\sigma < 8\ev$. The ionized reflection and smeared absorption models
both fail to describe the data below $0.5\kev$ when extended to lower
energies. Refitting the reflection model with the addition of an
absorption line at $0.54\kev$ to the full band data resulted in poor
fit ($\chi^2/dof = 1392.2/906$) with large emissivity index ($\beta =
9.17_{-0.03}^{+0.20}$) and small inner radius ($r_{in} < 1.239r_g$) as
before.  The smeared absorption model with the addition of the
absorption line provided the worst fit among the four models.

\section{Discussion}
We have studied the X-ray temporal and spectral characteristics of
Akn~564 and Mrk~1044. We found that the $0.2-2\kev$ and $2-10\kev$
band lightcurves are strongly correlated without any time delay.
However, the strength of this correlation decreases with increasing
separation between the soft and hard bands. We have found significant
correlation between the soft ($0.2-0.3\kev$, $0.2-0.5\kev$) and hard
($5-10\kev$, $4-10\kev$) band lightcurves at positive time delays. The
$4-10\kev$ band lags behind the $0.2-0.5\kev$ band by $1767\pm122\s$
(at $90\%$) in Akn~564. We note that Arevalo et al (2006) found time
delays that increase both with time scale and energy separation
between the energy bands. This means that there is no single, well
defined delay. Instead, the measured delay based on the
cross-correlation analysis is the averaged delay between the soft and
hard bands.  We find no correlation at zero time lag between the
$0.2-0.3\kev$ and $5-10\kev$ band emission from Mrk~1044.  We confirm
the presence of a break at $\sim 1.2\times 10^{-3}\hz$ in the power
density spectrum of Akn~564, earlier detected by Papadakis et al.
(2002) based on \asca{} observations. We also find that the PDS of
Akn~564 is energy dependent.  The hard band ($3-10\kev$) PDS,
consistent with a simple power law, is flatter and stronger than the
soft band ($0.2-0.5\kev$) PDS that shows clear evidence for the break.
These observations strongly suggest that different variability process
dominate the soft and hard band emission. This may suggest physically
distinct spectral components for the soft excess and the power-law
components.

\subsection{Implications of the observed time delays}
The observed hard band delays from Akn~564 demonstrate that the soft
excess emission cannot be the reprocessed emission of the primary X-ray
emission.  Fabian et al. (2002) presented the reflection dominated
ionized accretion disk model and successfully reproduced the strong
soft excess emission from the NLS1 galaxy 1H~0707-495. In this model,
the strong soft excess is caused by the multiple reflections of mostly
hidden hard X-ray continuum by the clumpy disk and the stronger
reflectivity of the partially ionized material below $\sim 0.7\kev$,
associated with an abrupt change in the opacity (Ross, Fabian \&
Ballantyne 2002; Fabian et al. 2004).  The soft X-ray excess emission
is the reprocessed emission in this model, hence it is unlikely that
the soft X-rays lead the hard X-ray emission.  This is, however, not a
strong constraint on the ionized reflection model. Many of the soft
X-ray lines expected from the accretion disk are excited by soft X-ray
photons with energies slightly greater than the line energies as the
photoelectric absorption cross section varies as $\sim E^{-3}$. The
ionizing soft X-ray photons can lead the hard ($3-10\kev$) band
emission due to the Compton delay of hard photons in the
Comptonization process. Since the life times of excited states,
responsible for the line photons, are very short compared to the
expected Compton delay between the soft and hard photons, the line
photons as well as the scattered soft X-ray emission can still lead
the hard band primary emission. However, in this case, the soft X-ray
excess emission is not expected to show rapid and large amplitude
variability as the reflection component and broad iron line are known
to show little variability in MCG-6-30-15 (Miniutti et al. 2007).
Moreover, the blurred ionized reflection model with physically
plausible parameters cannot reproduce the
observed soft X-ray excess emission from Akn~564 and Mrk~1044.  
     
Gierlinski \& Done (2004) presented an alternative model for the
origin of the soft excess emission.
In this model, the soft excess emission is mainly due to the strong
and smeared absorption near $\sim 0.7\kev$.  The soft excess and the
hard power law are part of the same and single continuum component
except for the small contribution from smeared emission lines.  This
means the soft and hard band emission arise from the same continuum
component or physical process - Comptonization. The time delay is a
natural outcome of the Comptonization process.  A general prediction
of Comptonization models is that the hard X-ray variations should lag
behind those in softer bands (Payne 1980).  Therefore it is natural to
attribute the observed time delays between hard and soft photons to
this process.  The time lag represents the difference in the photon
escape times between the soft and hard bands.  The hard band photons
tend to have undergone more scattering events, therefore travel longer
and escape the corona later than the soft band photons.  Thus the
observed time lag is consistent with a single continuum component and
hence with the smeared absorption model.  However, there are three
problems with the smeared wind model. ($i$) This model (XSPEC local
model {\tt swind1}; Gierli{\'n}ski \& Done 2004) does not describe the
X-ray spectrum of Akn~564 satisfactorily.  ($ii$) The $3-10\kev$ band
shows more variability power on short time scales than the
$0.2-0.5\kev$ band, which is not expected in the simple Comptonization
models in which the variability is caused by the variations in the seed
photons alone. ($iii$) The smearing velocity for Akn~564 suggested by the
{\tt swind1} model is very large ($v \sim 0.8c$) for disk winds.
Hydrodynamical simulations show that the line driven winds do not have
the required smearing velocity (N.  J. Schurch, private
communication).  It remains to be seen if the magnetically driven
winds can provide the required smearing.
 
The two component model, Comptonization in two separate regions, is a
viable option.  In this model, the soft excess emission arises from
the Comptonization in the low temperature (warm), optically thick
plasma while the hard power-law is produced by the Comptonization in
the hot plasma.  Indeed the EPIC-pn spectra of Akn~564 and Mrk~1044
are both well described by the two component model. The temperature of
the warm plasma is $\sim 175\ev$ for ~Akn~564 and $\sim135\ev$ for
Mrk~1044. This region, giving rise the soft excess emission, is
optically thick ($\tau \sim 45$). Physically, this component could be
the warm skin on the accretion disk surface or a region between the
disk and an optically thin inner flow (Magdziarz et al. 1998) or an
optically thick corona coupled with a truncated disk (Done \& Kubota
2006).  Depending on the geometry of the disk and the two component
corona system, there can be delay between the soft excess and high
energy power-law components.

However, the two component Comptonization model has difficulty in
explaining the constancy of the temperature of the Comptonizing region
producing the soft excess emission. Gierlinski \& Done (2004) derived
the temperature of the putative Comptonizing region in the range
$0.1-0.2\kev$ which is remarkably constant for 26 different PG
quasars. They concluded that the constancy of the temperature is
inconsistent with that expected from the span of the asymptotic
power-law index ($\Gamma_{thcomp} \sim 1.5-2.5$). However, the
$0.3-10\kev$ \xmm{} band only includes the high energy tail of the
optically thick Comptonized emission.  Both the peak of the soft
excess component ($kT\sim 0.1-0.2\kev$) and the asymptotic power-law
component fall below $0.3\kev$.  Using such data, it may be difficult
to measure accurately both the temperature and the asymptotic photon
index, particularly in the presence of line of sight neutral
absorption and complex warm absorber, both modifying the soft X-ray
spectrum below $1\kev$. NLS1 galaxies show a large diversity in their
soft X-ray \rosat{} photon indices ($\Gamma \sim 1.5-5.0$), suggesting
varying shape of the soft excess component (Boller, Brandt \& Fink
1996). The constancy of the temperature for the soft excess emission
is required to be investigated using a sample of NLS1 galaxies similar
to that used by Boller, Brandt \& Fink (1996).  Another possibility is
that the disk/corona geometry is different in NLS1 galaxies as they
are accreting close to the Eddington rate. In such situations, the
inner disk is likely thick and hotter. It remains to be seen if such
thick structure can have uniform temperatures across AGN with a large
range in their luminosity and black hole masses.

\subsection{The disk-corona geometry}

Inverse Compton scattering by thermal electrons provides the best
possible explanation for the observed smooth soft X-ray excess
emission from Akn~564 and Mrk~1044. Our spectral analysis favors two
physically distinct corona: high optical depth, low temperature corona
and high temperature, low optical depth corona. There are two possible
geometries with two component corona. The cool, optically thick corona
may be similar to the coupled disk corona geometry proposed by Done \&
Kubota (2006) for the very high state of black hole binaries. In this
geometry, the inner accretion disk is truncated and is coupled to an
optically thick corona, and gravitational energy release powers both
the disk and the optically-thick corona.  Comptonization of the disk
photons in this corona may give rise to the broad soft excess
component.  In addition to the optically thick corona, a hot corona is
responsible for the power-law component. The hot corona could be the
same as the patchy corona above the disk and is likely powered by
magnetic flares in the disk similar to the solar flares (Haardt,
Maraschi, \& Ghisellini 1994; Stern et al. 1995; Poutanen \& Svensson
1996; Poutanen \& Fabian 1999; Merloni \& Fabian 2001). The seed
photons could either be the soft photons from the disk or those
arising from the cool, optically thick corona.  Since the disk is
coupled to the optically thick corona, appreciable time lag between
the disk photons and the soft excess photons are not expected. The
variability of the disk flux and the soft excess emission is governed
by the variations in the accretion rate, while the variability of the
power-law component is governed both by the variation in the soft
photons as well as any variability process intrinsic to the corona
such as that related to variation in the electron population.  The
flatter power density spectrum of Akn~564 in the hard band may be
related to the additional variability of the power-law component due
to the changes in electron population on short time scales e.g.,
caused by magnetic flares above the accretion disk.

Another possible geometry is the ionized surface of a disk acting as
an optically thick and cool corona with a temperature of $\sim
0.1-0.3\kev$ (e.g, Magdziarz et al. 1998; see also O'Brian et al.
2001; Page et al. 2002; Vaughan et al. 2002). In this case, thermal
Comptonization of disk photons by thermal electrons in the hot surface
of the disk will produce the smooth soft excess. The steep power-law
extending to high energies can be produced in another purely hot
thermal/nonthermal or in a hybrid thermal-nonthermal plasma.

It is difficult to distinguish the two geometries described above. The
different power spectra of the soft and hard band X-ray emission may
constrain the two geometries. Recently McHardy et al. (2006) have
shown a strong correlation between the break frequency, black hole
mass and the bolometric luminosity for black hole binaries and AGN. If
the break frequencies are related to the characteristic radii from the
central source, their result suggests a characteristic size that
appears to decrease with increasing accretion rate or luminosity. The
clear evidence for a break in the soft band PDS of Akn~564 suggests a
characteristic size associated with the soft X-ray excess emission
which can be the size of the optically thick corona or the radius of
the inner truncated disk in the model suggested by Done \& Kubota
(2006). It is also possible that the disk surface below an inner
radius is ionized and the temperature of the inner disk surface is
higher than that expected from the standard disk.  The inner disk
surface may then be responsible for the soft excess emission. The lack
of a clear break in the hard band PDS of Akn~564 may suggest that
power-law emission is probably not associated with a characteristic
size and it arises from an extended corona similar to a patchy corona
above an accretion disk.  Future long observations of Akn~564 are
required to clearly demonstrate the presence or absence of a break in
the hard band PDS.

\section{Conclusions}
We have presented temporal and spectral study of \xmm{} observations
of Akn~564 and Mrk~1044. The main results are as follows:
\begin{enumerate}
\item The $0.2-2\kev$ and $2-10\kev$ band X-ray emission from both the
  NLS1 galaxies are strongly correlated without any time delay.
  However, the variations in the $0.2-0.5\kev$ band are found to lead
  those in the $4-10\kev$ band by $1768\pm122\s$ ($90\%$) in Akn~564.
  We do not find any correlation between the $0.2-0.3\kev$ and
  $5-10\kev$ X-ray emission from Mrk~1044 at zero time delay. The two
  bands appear to anti-correlate at a positive lag of $\sim 1000\s$
  and likely correlated at a lag of $\sim3000 - 5000\s$.  Long
  observations of Mrk~1044 are required to confirm the time delay.
\item The full band power density spectrum of Akn~564 has a break at
  $(1.2\pm0.3)\times 10^{-3}\hz$, corresponding to a time scale of
  $\sim 1000\s$, similar to the delay between $0.2-0.5$ and $4-10\kev$
  bands. It is not clear if there is any physical relation between the
  two time scales. There is a clear evidence of a break in the soft
  band PDS but the hard band does not show a clear evidence for the
  break.
\item The soft ($0.2-0.5\kev$) band PDS of Akn~564 is significantly
  steeper and weaker than the hard ($3-10\kev$) band PDS. This implies
  that the power-law component is more variable than the soft excess
  component on shorter time scales. The implication of this result on
  the Comptonization models is that the variations in the power-law
  component are not only caused by the variations in the seed photons
  but also by the variations in the hot electron population. Thus
  electron injection on short time scales ($\la 1000\s$) are required.
  This could be evidence for magnetic flares thought to power the
  corona above the accretion disk.
\item The soft excess emission from Mrk~1044 is featureless in the
  EPIC-pn data and is well described by the smeared absorption or
  optically thick Comptonization. The RGS data show evidence for warm
  absorbers in Akn~564. The EPIC-pn spectrum of Akn~564 is
  well described by  a complex model consisting of optically thick 
  thermal Comptonization in a low temperature ($\sim 0.15\kev$) plasma
  and a steep power-law, modified by two phase warm absorber medium
  and the Galactic absorption. The ionized reflection and smeared wind
  models fail to describe the data satisfactorily. 
\item The temporal and spectral characteristics of Akn~564 and
  Mrk~1044  are consistent with a two component corona -- a compact low
  temperature, optically thick corona and an extended hot corona. The
  compact corona could be an inner optically thick region coupled to a
  truncated disk or the ionized surface of an inner untruncated disk.
  The soft excess emission can be produced in the inner, optically
  thick corona and the power-law component is produced in the hot
  corona.
\end{enumerate}

\acknowledgements We are grateful to the referee, I.  Papadakis, for
his detailed comments and suggestions that improved this paper
significantly. We thank M. Nowak for writing a number of useful ISIS
functions that have been used in this paper. GCD gratefully
ackowledges the support of NASA grants NNX07AE99G and NNX06AE38G. This
work is based on observations obtained with \xmm{}, an ESA science
mission with instruments and contributions directly funded by ESA
Member States and the USA (NASA). This research has made use of data
obtained through the High Energy Astrophysics Science Archive Research
Center Online Service, provided by the NASA/Goddard Space Flight
Center.

\clearpage 

\clearpage

\clearpage


\clearpage

\end{document}